\documentclass[11pt]{article} 
\usepackage{hyperref}
\pdfoutput=1 

\begin{document} 

\title{The 3D structure of turbulent channels up to $\mathrm{Re}_\tau=4000$} 
\author{Adri\'an Lozano-Dur\'an \& Javier Jim\'enez\\ 
\\\vspace{6pt} Computational Fluid Mechanics Lab., \\ Universidad Polit\'ecnica de Madrid, Spain} 
\maketitle 
\begin{abstract} 
\noindent Several time-resolved fluid dynamics videos of turbulent channels from $\mathrm{Re}_\tau=180$ to 
$\mathrm{Re}_\tau=4000$ are presented. The videos show the temporal evolution of sweeps (bluish) 
and ejections (reddish) in one half of the channel (only the bottom wall is shown). 
The color changes from dark for points close to the wall, to bright for those reaching
the center of the channel. As the Reynolds number increases the scale separation becomes 
more clear and the complexity of the dynamics observed rises.

\end{abstract} 
\section{Introduction} 

\noindent The efforts to describe wall-bounded turbulent flows in terms of coherent motions date at 
least to the experiments in \cite{kim:kli:rey:71}. These structures have played an important role in the 
understanding of turbulence organization and its dynamics. Data from Direct Numerical Simulations 
allows us to study the properties of these objects in three dimensions for different snapshots, 
but their dynamics can not be completely solved without tracking them in time. That temporal 
evolution has already been studied for either very small or large structures at moderate Reynolds 
numbers but a temporal analysis of 3D structures spanning from the smallest to the largest scales 
across the logarithmic layer has not been performed yet. Some attempts have been done \cite{loji2011}, 
although in very small domains and at moderate Reynolds numbers. 
\\
\noindent In the present video, 3D sweeps and ejections are plotted using turbulent channels datasets performed 
in a physical domain big enough to not influence the largest structures in the logarithmic region 
(for the highest Reynolds numbers)
and small enough to obtain a reasonable and tractable amount of data. The highest Reynolds number 
ensures the presence of a wide range of scales and a healthy logarithmic region, 
that would allow us to study the dynamics of the coherent structures in time.

\section{Results shown in the video} 

\noindent The flow is moving on average in the streamwise direction $x$, $z$ is the spanwise 
direction and $y$ the wall-normal one. The superindex $+$ denotes wall units and the 
wall is located at $y^+=0$. The domain size in all the cases is $L_x=2\pi h$,
$L_z=\pi h$ and $L_y=2 h$ where $h$ is the channel half height.
The incompressible flow is integrated in
the form of evolution equations for the wall-normal vorticity and for the
Laplacian of the wall-normal velocity, as in \cite{kim:moi:mos:87}, and the
spatial discretization is dealiased Fourier in the two wall-parallel
directions. Reynolds numbers $\mathrm{Re}_\tau=180$ and $950$ use Chebychev 
polynomials in $y$ whereas $\mathrm{Re}_\tau=2000$ and $4000$ use
seven-point compact finite differences. Time stepping is the third-order
semi-implicit Runge-Kutta in \cite{spalart91}.
\\
\noindent The coherent structures represented are characterized in \cite{lo:flo:jim:2012}.
They are the structures contributing most to the Reynolds stresses and are obtained 
by extending the one-dimensional quadrant analysis of \cite{lu:wil:73} to three dimensions.
They are connected regions satisfying
\begin{equation}
\tau({\bf x}) > H {u'(y) v'(y)},
\end{equation}
where $\tau({\bf x })=-u({\bf x })v({\bf x})$ is the
instantaneous point-wise tangential Reynolds stress with $u$ and $v$ are the 
velocity fluctuations in streamwise and wall-normal directions respectively.
$H$ is the hyperbolic-hole size and it is set to $1.75$ in all the cases.
An object is classified as belonging to the different
quadrants according to the signs of the mean $u$ and $v$.
\\
\noindent Only structures beloging to quadrants 2 (ejections) and 4 (sweeps) are plotted.
The sweeps are colored in blue and ejections in red in one half of the channel. 
Their color changes from dark for points close to the wall, to bright for those reaching
the center of the channel.
\\

\newpage

\subsection*{Acknowledgements}
This work was supported in part by CICYT, under grant TRA2009-11498, and by
the European Research Council, under grant ERC-2010.AdG-20100224. A.
Lozano--Dur\'an was supported by an FPI fellowship from the Spanish
Ministry of Education and Science. The computations were made possible 
by generous grants of computer time from CeSViMa (Centro de Supercomputaci\'on y 
Visualizaci\'on de Madrid) and BSC (Barcelona Supercomputing Centre). 

\bibliographystyle{plain} 
\bibliography{mybib}      

\end{document}